\newcounter{myctr}
\def\myitem{\refstepcounter{myctr}\bibfont\noindent\ifnum\themyctr>9
\else\phantom{0}\fi\hangindent17pt\themyctr.\enskip}

\documentclass[final]{ws-ijqi}
\usepackage{graphicx,bbm,amsmath,color}
\newcommand{\real}[1]{\Re\mbox{e}[#1]}
\newcommand{\immag}[1]{\Im\mbox{m}[#1]}
\newcommand{\bmsigma}{\boldsymbol \sigma} 
\newcommand{\bmLambda}{\boldsymbol \Lambda} \def\D{{\rm D}}
\newcommand{\bmlambda}{\boldsymbol \lambda} \def\X{\boldsymbol{X}}
\def\Tr{\hbox{Tr}} 

\begin{document}
\markboth{Marco G. Genoni and Matteo G. A. Paris}{Purity 
and non-Gaussianity in truncated Hilbert spaces}
%%%%%%%%%%%%%%%%%%%%% Publisher's Area please ignore %%%%%%%%%%%%%%
\catchline{}{}{}{}{}
%%%%%%%%%%%%%%%%%%%%%%%%%%%%%%%%%%%%%%%%%%%%%%%%%%%%%%%%%%%%%%%%%%%
\title{NON-GAUSSIANITY AND PURITY IN FINITE DIMENSION}
\author{MARCO G. GENONI$^{1,2}$, and MATTEO G. A. PARIS$^{1,2,3}$}
\address{${}^1$Dipartimento di Fisica dell'Universit\`a di Milano,
I-20133, Milano, Italia.\\
${}^2$CNISM, UdR Milano Universit\`a, 20133, Milano, Italia \\
${}^3$ISI Foundation, I-10133, Torino, Italia. }
%\date{\today}
\maketitle
\begin{abstract}
We address truncated states of continuous variable systems and analyze
their statistical properties numerically by generating random states 
in finite-dimensional Hilbert spaces.  In particular, we focus to the
distribution of purity and non-Gaussianity for dimension up to 
$d=21$. We found that both quantities are
distributed around typical values with variances that decrease for
increasing dimension. Approximate formulas for typical purity and
non-Gaussianity as a function of the dimension are derived.
\end{abstract}
%%%%%%%%%%%%%%%%%%%%%%%%%%%%%%%%%%%%%%%%%%%%%%%%%%%%%%%%%%%%%%%%%%%
%\keywords{purity; non-Gaussianity; random quantum states}
%%%%%%%%%%%%%%%%%%%%%%%%%%%%%%%%%%%%%%%%%%%%%%%%%%%%%%%%%%%%%%%%%%%
\section{Introduction}
Quantum states of continuous variable (CV) systems with bounded
occupation number $N$ correspond to finite superpositions, or
mixture, of Fock states $\varrho_N=\sum_{nk=0}^N\varrho_{nk}
|n\rangle\langle k|$ and are usually referred to as truncated states.
Truncated states may be obtained by heralding techniques from entangled
sources \cite{r1,r2} or by quantum state engineering in cavity QED
\cite{q1} and nonlinear interferometry \cite{t1}.
Random quantum states in finite-dimensional Hilbert spaces have been
widely investigated, mostly to find typical values of nonlinear
functions of the density matrix which are relevant for quantum
information processing.  In particular, the distribution of 
entanglement for bipartite states and the volume of separable states
\cite{zyc01,zyc02,cap01} have been examined. Here we  
address states in Hilbert spaces with finite dimension $d$ as truncated 
continuous variables states with maximum occupation number $N=d-1$. 
Our goal is to characterize their statistical properties with focus to 
state purity and non-Gaussianity (nonG) and, in particular, 
to find their typical values by random generation of states in
finite dimensional Hilbert spaces.
\par
Gaussian states play a relevant role in CV quantum information
processing \cite{GRG}. In particular, teleportation, cloning and dense
coding have been implemented by using Gaussian states and Gaussian
operations. On the other hand, it has been demonstrated that the
Gaussian sector of the Hilbert space is not enough to perform long
distance quantum communication: protocols as entanglement distillation
and entanglement swapping need nonG maps, thus yielding nonG states.
Moreover, by using nonG states and operations, teleportation
\cite{Tom,IPS2a,IPS2b} and cloning \cite{nonGclon} of quantum states may
be improved. De-Gaussification protocols for single-mode and two-mode
states have been indeed proposed \cite{Tom,IPS2a,IPS2b,IPS1,KorolkovaKerr} 
and realized \cite{IPS_Wenger}.
\par
In this sense, the nonG character of states and operations represents a
resource for CV quantum information and a question arises on whether the
nonG character is a general feature of truncated states.  In order to
gain information about nonG properties of finite-dimensional states we
exploit a recently proposed nonG measure \cite{nonG_HS}.  We
then generate uniformly random quantum states for different dimension of
the truncated Hilbert space \cite{zyc01} and analyze the distribution of
nonG and purity of these states and their average values.
We focus on the dependence on the dimension of the Hilbert space and use
our results to draw some conjectures about the behaviour in higher
dimensions.
\par 
The paper is structured as follows. In the next section we briefly
review the generation of uniformly distributed quantum states in a finite
dimensional Hilbert space. In Section \ref{s:Gauss} we review the basic
properties of Gaussian states and of the measure of the nonG character.
Then, in Section \ref{s:results} we evaluate the typical values and the
distributions of nonG and purity for random quantum states in
finite dimensional Hilbert spaces. Section \ref{s:out} closes the paper
with some concluding remarks.
%%%%%%%%%%%%%%%%%%%
\section{Random quantum states} \label{s:random}
In order to generate states randomly distributed in a 
$d$-dimensional Hilbert we consider the spectral decomposition
$\varrho = \sum_{n=0}^{d-1} \lambda_n P_n$, $\sum_{n=0}^{d-1}
\lambda_n=1$, $\lambda_n \geq 0$
where $P_n$ form a complete set of orthogonal projectors. Therefore, 
we may view the set of quantum states as the Cartesian product \cite{zyc01}, 
$\mathcal{S} = \mathcal{P}\times \Delta$ 
where $\mathcal{P}$ denotes the family of complete
sets of orthonormal projectors and where $\Delta$ denotes
the simplex, \emph{i.e.} the subset of the
$(d-1)$-dimensional linear sub-manifold of real space $\mathbb{R}^N$,
defined by the trace condition $\sum_{n=0}^{d-1}\lambda_n=1$. This 
representation of quantum states in $d$-dimensional Hilbert spaces
corresponds to the decomposition $\varrho = U D U^{\dag}$, 
where $U$ denotes a unitary matrix and $D$ a diagonal matrix with trace
equal to one. A  uniform distribution of density matrices may be thus 
obtained by choosing the uniform distribution 
on the group of unitary transformations $U(N)$ (Haar measure) and on the
set of diagonal matrices $D$, {\em i.e.} the distribution on the simplex. 
Following these lines we have generated random quantum states upon
employing an algorithm to generate random $U(N)$ matrices according to 
the Haar measure \cite{zyc03} as well as an algorithm to generate 
random points on the simplex \cite{zyc01}.
\section{Gaussian states} \label{s:Gauss}
In this section we will review the definition and the principal 
properties of Gaussian states, by using the quantum optical terminology
of modes carrying photons, though our theory applies to general
bosonic systems. Let us consider a CV systems of $n$ modes described
by the mode operators  $a_k$, $k=1\dots n$, satisfying the commutation relations
$[a_k,a_j^{\dag}]=\delta_{kj}$. A quantum state $\varrho$ of the
$n$ modes is fully described by its characteristic function
\cite{Glauber2}
$
\chi[\varrho](\bmlambda) = \Tr[\varrho\,D(\bmlambda)]
$
where $D(\bmlambda) = \bigotimes_{k=1}^n D_k(\lambda_k)$
is the $n$-mode displacement operator, with $\bmlambda =
(\lambda_1,\dots,\lambda_n)^T$, $\lambda_k \in \mathbbm{C}$, and 
$
D_k(\lambda_k) =\exp\{\lambda_k a_k^{\dag} - \lambda_k^* a_k \}
$
is the single-mode displacement operator.
The canonical operators are 
$q_k = \frac{1}{\sqrt{2}}(a_k + a^{\dag}_k)$, 
$p_k = \frac{1}{i\sqrt{2}}(a_k - a_k^{\dag})$ 
with commutation relations given by $[q_j,p_k]=i\delta_{jk}$.
Upon introducing the  real vector $\boldsymbol{R}=(q_1,p_1,\dots,q_n,p_n)^T$,
we define vector of mean values $\X=\X[\varrho]$ and the covariance
matrix $\bmsigma=\bmsigma[\varrho]$ as 
${X}_j = \langle R_j \rangle$ and $
\sigma_{kj} = \frac{1}{2}\langle \{ R_k,R_j \}\rangle - \langle R_j
\rangle\langle R_k\rangle$, where $\{ A,B \} = AB + BA$ denotes the anti-commutator, and
$\langle O \rangle = \Tr[\varrho\:O]$ is the expectation value
of the operator $O$.
A quantum state $\varrho_G$ is referred to as a Gaussian state if
its characteristic function has the Gaussian form
$
\chi[\varrho_G](\bmLambda) = \exp \left\{ - \frac{1}{2}
\bmLambda^T \bmsigma \bmLambda + \X^T \boldsymbol{\Omega} \bmLambda \right\}
$
where $\bmLambda$ is the real vector $\bmLambda = (\hbox{Re}
\lambda_1, \hbox{Im}\lambda_1, \dots, \hbox{Re} \lambda_n,
\hbox{Im}\lambda_n)^T$. Of course, once the covariance matrix and
the vector of mean values are given, a Gaussian state is fully
determined. For a single-mode system the most general
Gaussian state can be written as
$
\tau= D(\alpha) S(\zeta)
\nu(n_t) S^\dag (\zeta) D^\dag(\alpha) ,
$
%%MG: ho cambiato \varrho_G in \tau$
$D(\alpha)$ being the
displacement operator, $S(\zeta) = \exp[\frac12 \zeta (a^{\dag})^2
- \frac12 \zeta^* a^2]$ the squeezing operator, $\alpha,\zeta \in
{\mathbbm C}$, and $\nu(n_t)=(1+n_t)^{-1} [n_t/(1+n_t)]^{a^\dag
a}$ a thermal state with $n_t$ average number of photons. 
Its matrix elements in the Fock basis are given by \cite{Marian}
%%MG: ho cambiato "The matrix elements" in "Its matrix elements"
\begin{align}
\langle l|\tau|m\rangle &= \frac{K}{(l!m!)^{1/2}} \sum_{k=0}^{\min[l,m]}
k! \binom{l}{k}\binom{m}{k} \tilde{A}^k (\frac{1}{2}\tilde{B})^{(l-k)/2}
(\frac{1}{2}\tilde{B}^{*})^{(m-k)/2} \nonumber \\
&\qquad \times H_{l-k}((2\tilde{B})^{-1/2}
\tilde{C})
H_{m-k}((2\tilde{B}^{*})^{-1/2}
\tilde{C}^{*}) \label{eq:GaussFock}
\end{align}
where 
\begin{align}
\tilde{A} &= \frac{A(1+A) - |B|^2 }{(1+A)^2 - |B|^2 } \quad 
\tilde{B} = \frac{C}{(1+A)^2 - |B|^2 } \quad 
\tilde{C} = \frac{(1+A)C + BC^{*}}{(1+A)^2 - |B|^2 } \nonumber \\
K &= [(1+A)^2 -|B|^2]^{-1/2} 
%\nonumber \\ &\qquad 
\exp \left\{ - \frac{(1+A)|C|^2 +\frac{1}{2}[B(C^*)^2 + B^*C^2]}
{(1+A)^2 - |B|^2} \right\} \nonumber
\end{align}
$H_n(x)$ denotes a Hermite polynomial and 
\begin{align}
A = \frac{ \sigma_{11}+\sigma_{22} -1}{2} \quad 
C = \frac{ X_1 + i X_2}{\sqrt{2}} \quad %\nonumber \\ 
\real{B} = \frac{ \sigma_{22} - \sigma_{11}}{2} \quad 
\immag{B} = -\sigma_{12} \nonumber 
\end{align}
%%%%%%%%%%%%%%%%%%%%
\subsection{A measure of non-Gaussianity}
The non-Gaussian character of a quantum state
$\varrho$ may be quantified as the squared Hilbert
distance between $\varrho$ and a reference Gaussian state $\tau$, 
normalized by the purity of $\varrho$ itself, in formula \cite{nonG_HS}
\begin{align}
\delta[\varrho] &= \frac{ \D_{HS}^2[\varrho,\tau]}{\mu[\varrho]} 
\label{eq:nonG1} 
\end{align}
where $\D_{HS}[\varrho,\tau]$ denotes the Hilbert-Schmidt distance 
between $\varrho$ and $\tau$, {\em i.e.} $\D_{HS}^2[\varrho,\tau] 
= \frac{1}{2} \Tr[(\varrho - \tau)^2] = \frac12 (\mu[\varrho] + \mu[\tau] 
- 2\kappa[\varrho,\tau])$
with $\mu[\varrho]=\hbox{Tr}[\varrho^2]$ and
$\kappa[\varrho,\tau]=\hbox{Tr}[\varrho\:\tau]$
denoting the purity of $\varrho$ and the overlap between
$\varrho$ and $\tau$ respectively. The Gaussian reference $\tau$
is chosen as the Gaussian state with the same covariance matrix 
$\bmsigma$ and the same vector $\X$ of $\varrho$, that is
$\X[\varrho] = \X[\tau]$ and $\boldsymbol{\sigma}[\varrho] = 
\boldsymbol{\sigma}[\tau]$. The nonG measure $\delta[\varrho]$
vanishes iff $\varrho$ is a Gaussian state, it is invariant under 
symplectic transformations and have been employed 
to analyze the evolution of quantum states undergoing 
Gaussification and de-Gaussification protocols \cite{nonG_HS}.
%%%%%%%%%%%%%%%%%%%%%%
\section{Non-Gaussianity and purity of random quantum states}
\label{s:results}
\begin{figure}[h]
\centerline{
\includegraphics[width=0.24\textwidth]{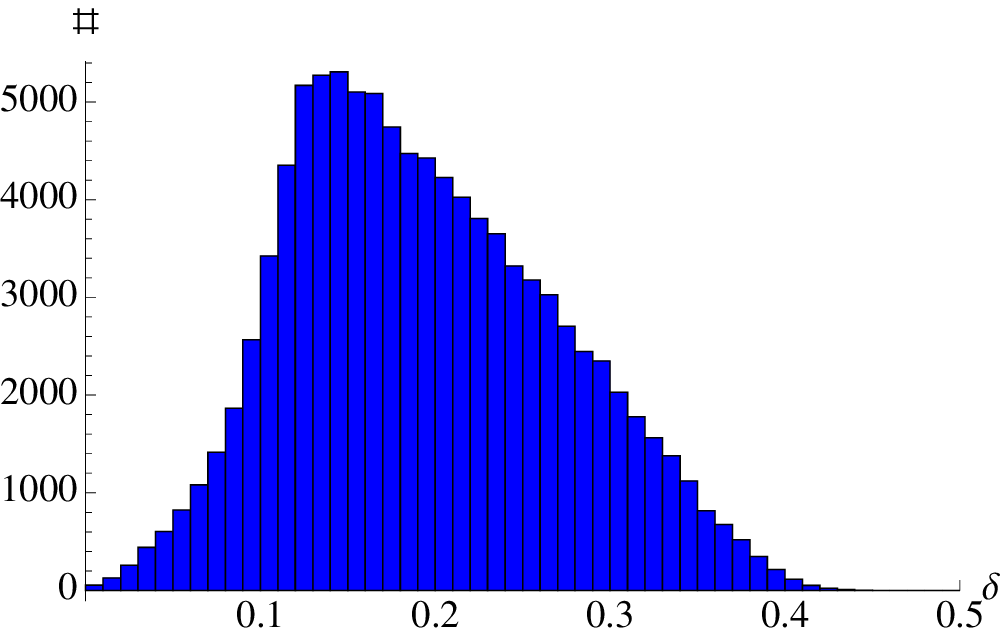}
\includegraphics[width=0.24\textwidth]{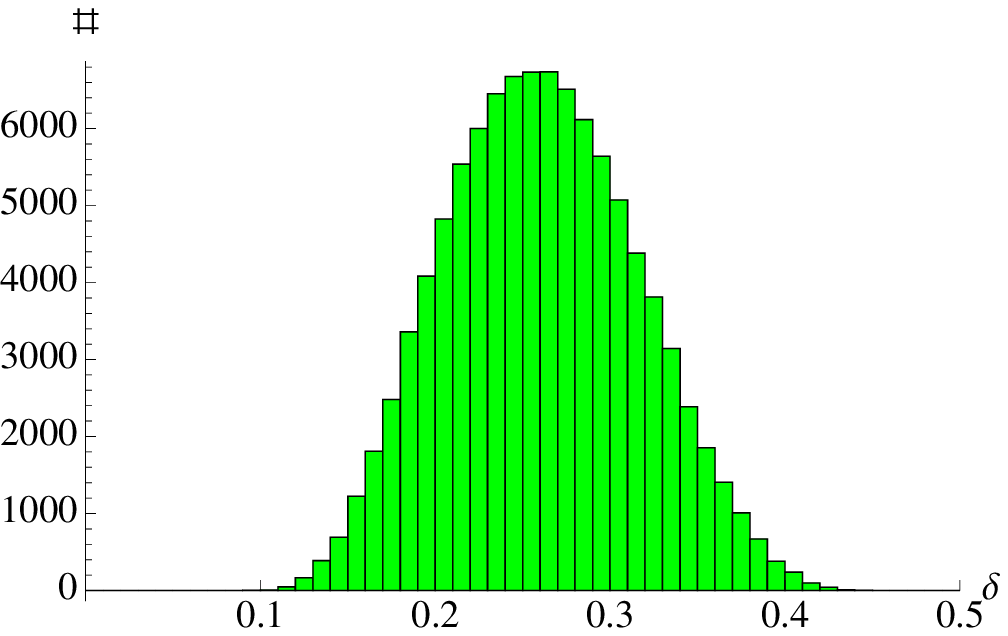}
%}
%\vspace*{4pt}
%\centerline{
\includegraphics[width=0.24\textwidth]{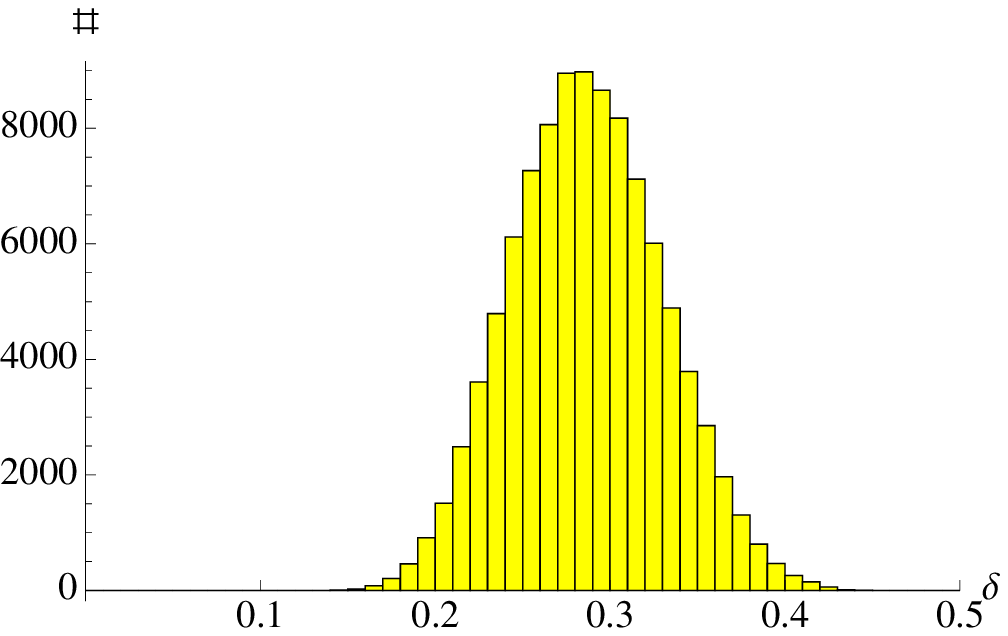}
\includegraphics[width=0.24\textwidth]{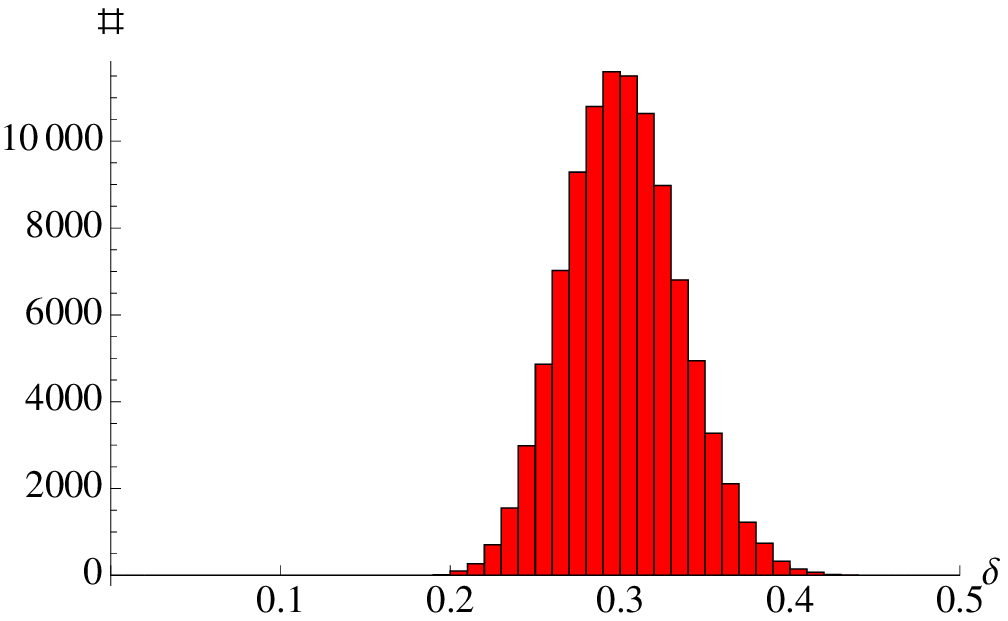}}
%\centerline{\psfig{file=hist_d.eps}} 
%\vspace*{8pt}
\caption{Histograms corresponding to the distributions
of nonG of $10^5$ random quantum states for different
dimensions of the Hilbert space. From left to right: 
$N=2$, $N=5$, $N=10$, $N=20$. \label{f:hist_d}}
\end{figure}

\begin{figure}[h]
\centerline{
\includegraphics[width=0.4\textwidth]{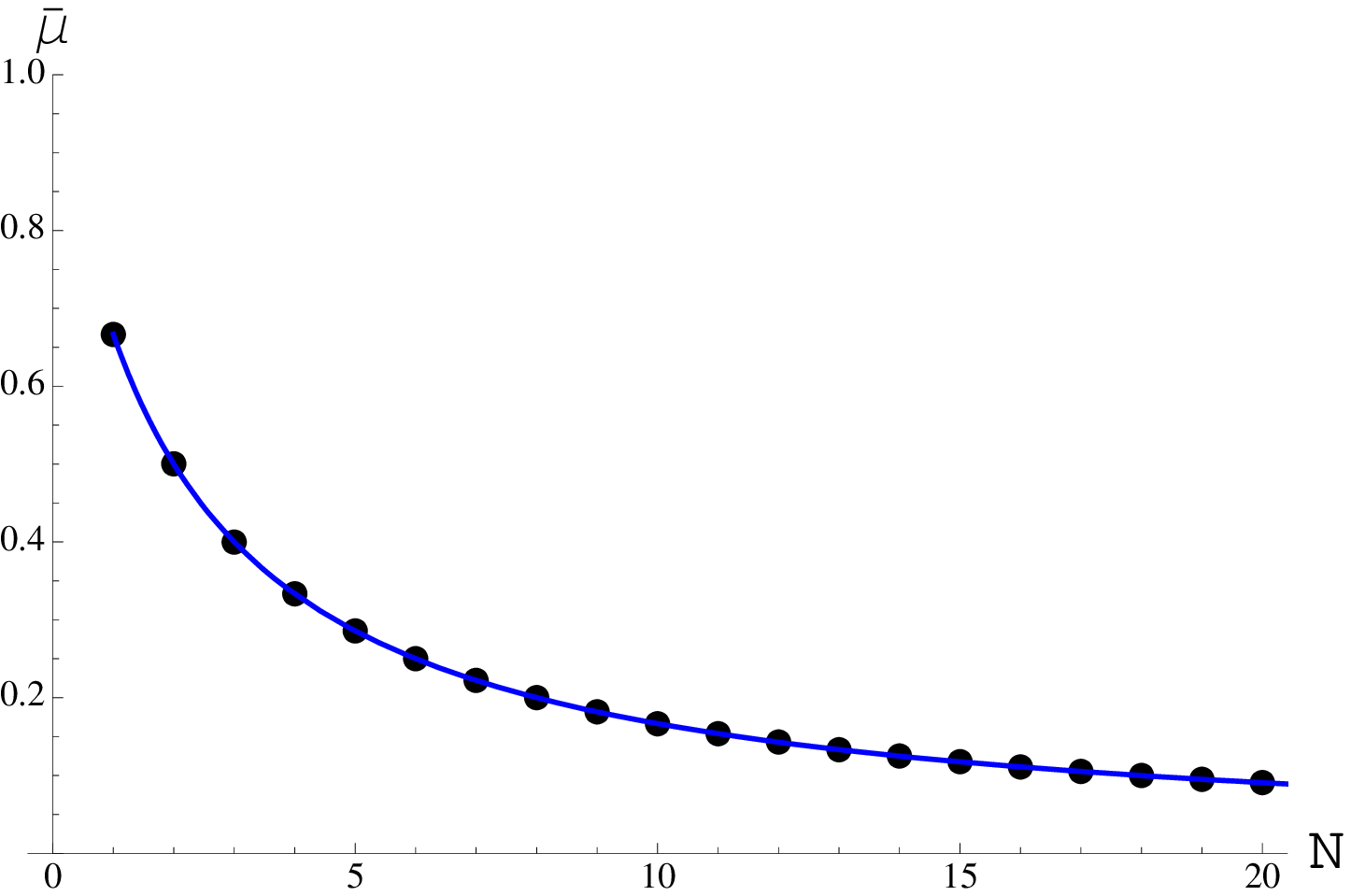}
\includegraphics[width=0.4\textwidth]{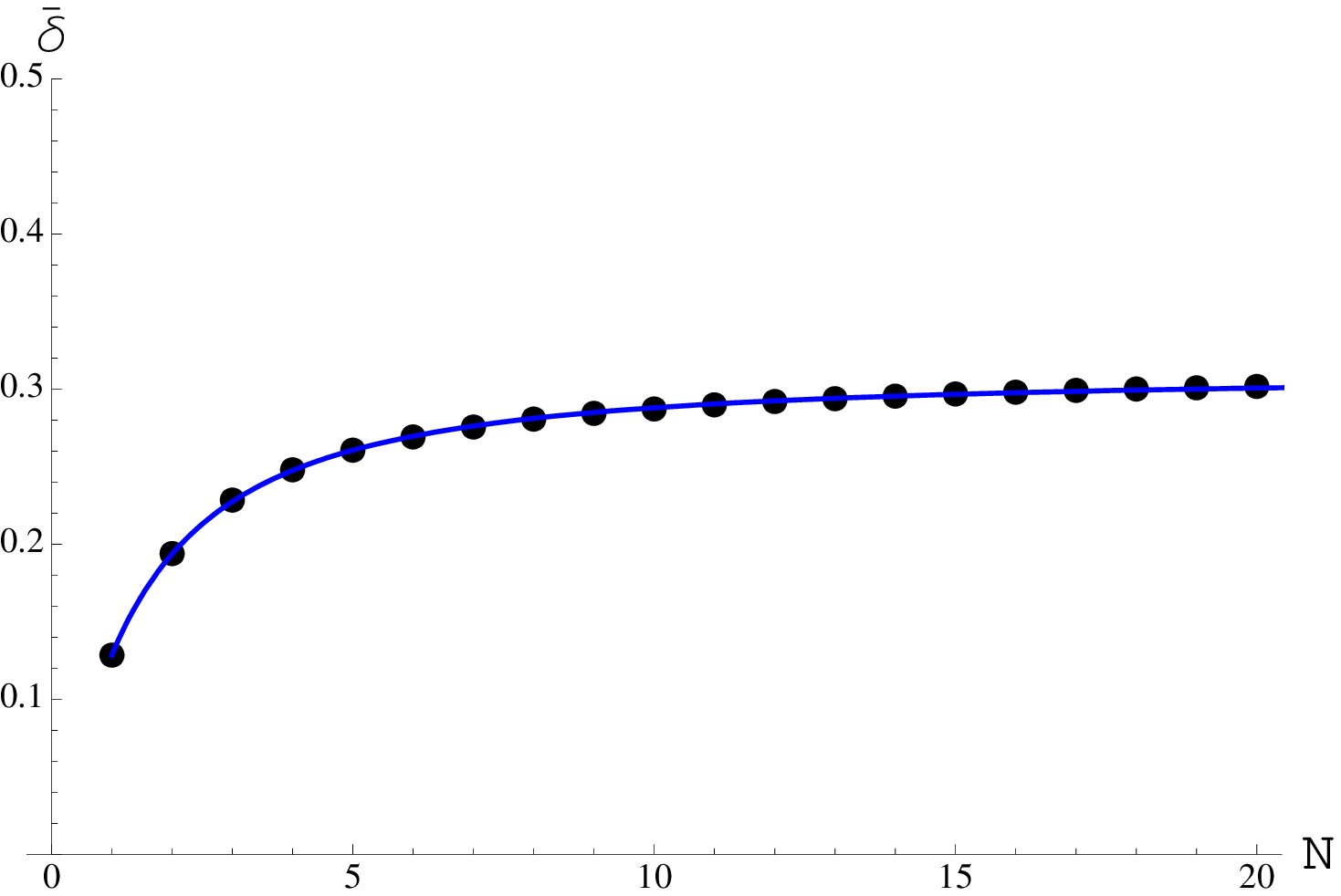}}
%\centerline{\psfig{file=hist_mu02.eps}\psfig{file=hist_mu05.eps}} 
\caption{
(Left) Black points:
typical purity as a function of the maximum occupation number $N$. 
Blue line: fitting function
$\overline{\mu}_f$ for the typical purity as a function of $N$.
(Right) Black points:
typical nonG as a function of the maximum occupation number $N$. 
Blue line: fitting function 
$\overline{\delta}_f$ for the typical nonG as a function of $N$.
\label{f:fit}}
\end{figure}

\begin{figure}[h]
\centerline{
\includegraphics[width=0.4\textwidth]{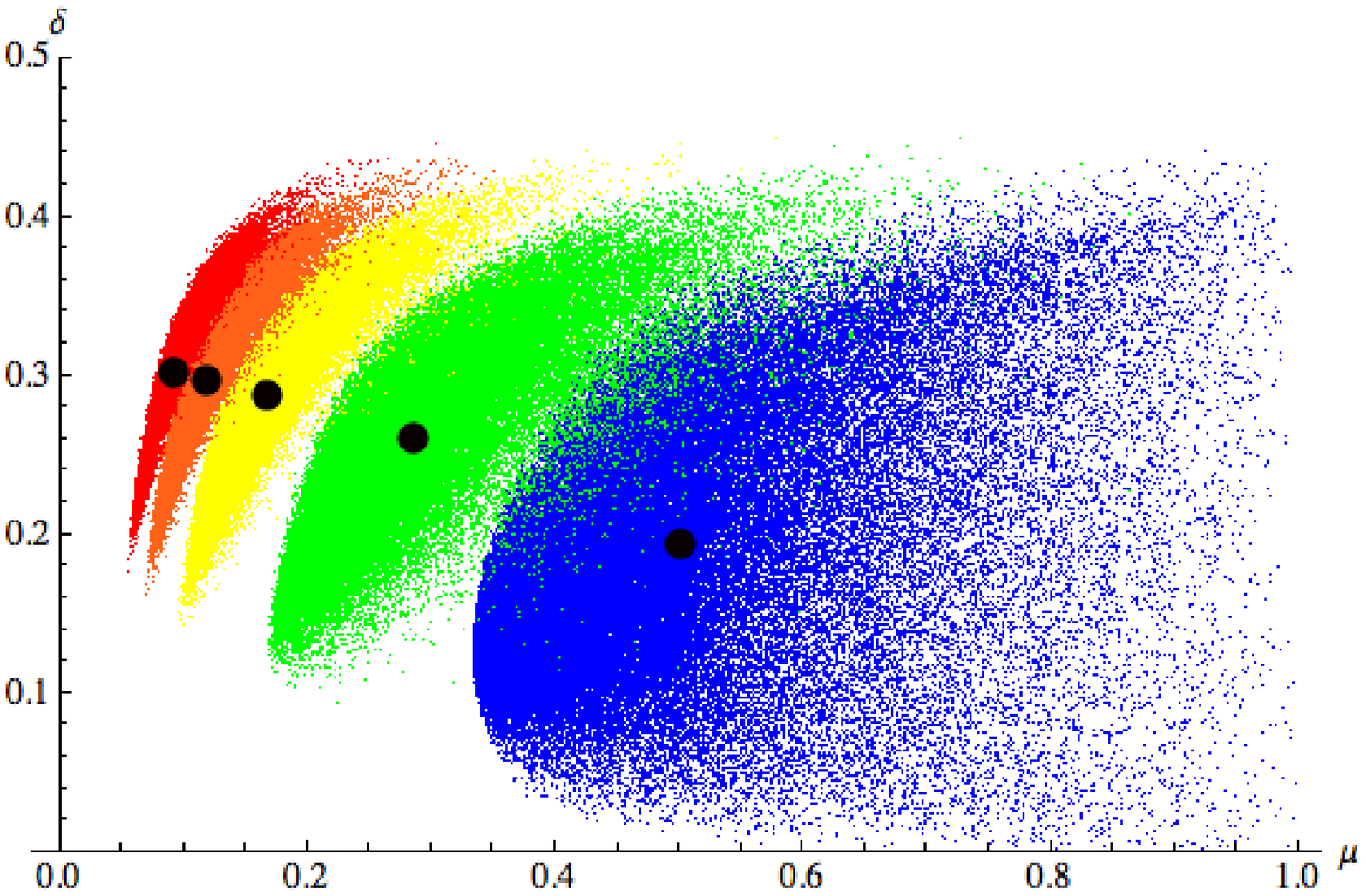}$\quad$
\includegraphics[width=0.4\textwidth]{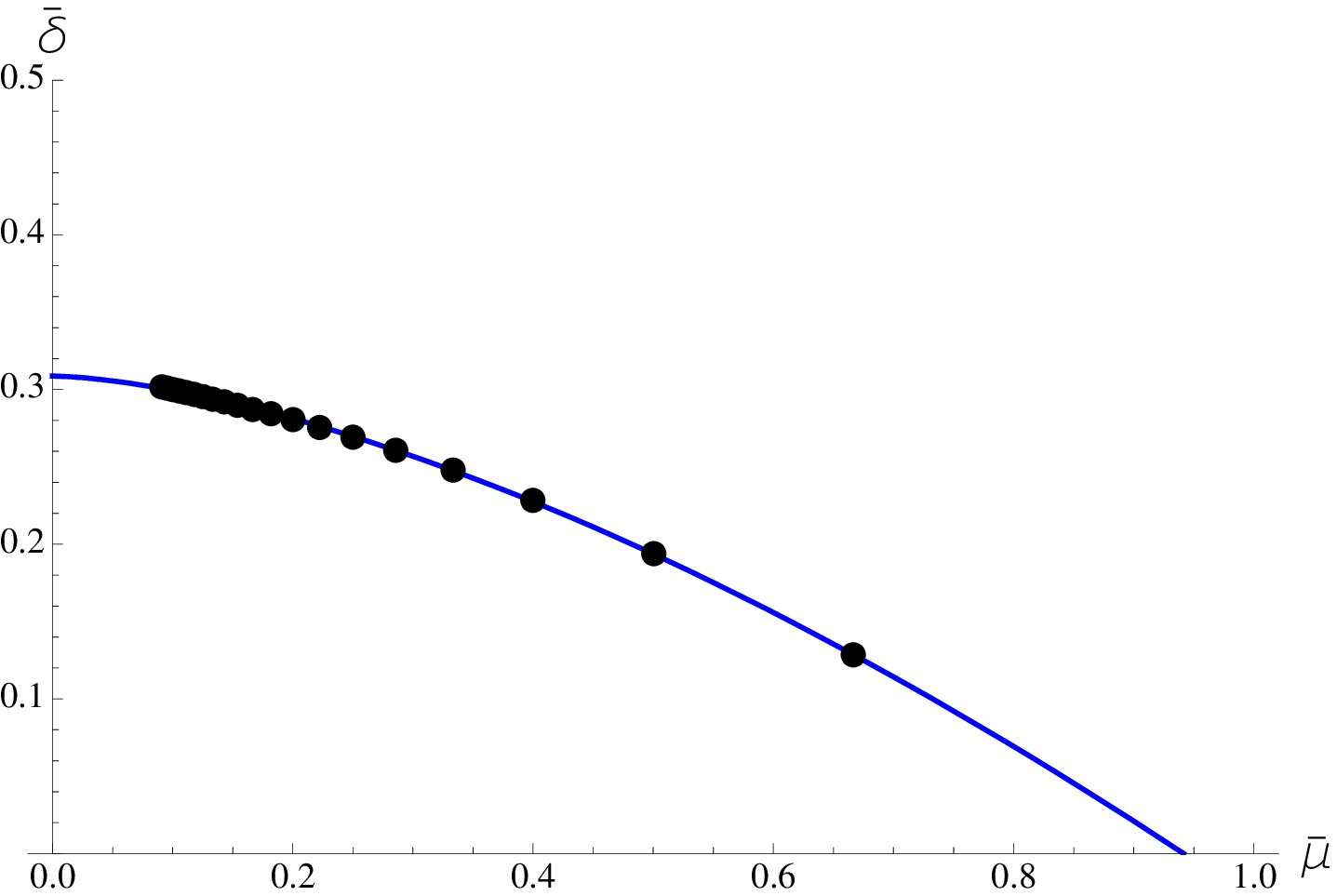}
}
\caption{
Left: Purity and nonG of random states as points in the plane 
$(\mu,\delta)$. Different colors correspond to 
different dimension of the Hilbert space. The black points 
correspond to the average (typical) purity and nonG at each dimension.
\textcolor{blue}{blue}: $N=2$.
\textcolor{green}{green}: $N=5$.
\textcolor{yellow}{yellow}: $N=10$.
orange $N=15$.
\textcolor{red}{red}: $N=20$.
Right: Typical purity and nonG in the plane
$(\mu,\delta)$ varying the dimension $N=2,\dots,20$.
The blue line correspond to the approximate formula reported
in the text.\label{f:pur_nong}}
\end{figure}

We have generated $10^5$ random quantum states
 $\varrho_N=\sum_{nk=0}^N\varrho_{nk} |n\rangle\langle k|$ 
%\begin{align}
%\varrho_{\hbox{\scriptsize N}} = \sum_{n,k=0}^N \varrho_{n,k} |n\rangle\langle k| 
%\end{align}
in finite dimensional subspaces, 
$\mathrm{dim}(H)=N+1$ 
($N=\{1,\dots,20\}$), following the 
algorithm explained in Section \ref{s:random}. 
We have evaluated the vector of mean values
$\X$ and the covariance matrix $\bmsigma$ for each generated state $\varrho_N$,
the corresponding reference Gaussian state $\tau$, as well as
 parameters $A$, $B$ and $C$. 
Then, using Eq. (\ref{eq:GaussFock}) we have reconstructed
 the density matrix elements of $\tau$, truncating the Hilbert space
 upon checking the normalization condition
$\Tr[\tau]=1$ up to an error of $10^{-4}$. We have evaluated 
the purity of the state $\mu[\varrho]$, its 
nonG $\delta[\varrho]$ and its symplectic eigenvalue $s[\varrho]$.
The corresponding average values along with the standard deviations are
reported in Table \ref{t:avg}, where we also report the average values
and the standard deviations of the purity of the reference 
Gaussian state $\mu[\tau]$ and of the overlap $\kappa[\varrho,\tau]$.
\begin{table}[h]
\caption{Average values and standard deviations of state purity 
$\mu[\varrho]$, purity of the Gaussian reference $\mu[\tau]$, 
overlap $\kappa[\varrho,\tau]$, nonG $\delta[\varrho]$,
symplectic eigenvalue $s[\varrho]$ evaluated on $10^5$ random quantum 
states for different dimensions $d=N+1$.}
\label{t:avg}
\centerline{{\tablefont 
\begin{tabular}{c|c @{$\pm$} c |c @{$\pm$} c |c @{$\pm$} c
|c @{$\pm$} c |c @{$\pm$} c}
$N$ & \multicolumn{2}{c|}{$\overline{\mu[\varrho]}_{\hbox{\scriptsize N}}$} &
\multicolumn{2}{c|}{$\overline{\mu[\tau]}_{\hbox{\scriptsize N}}$} &
\multicolumn{2}{c|}{$\overline{\kappa[\varrho ,\tau ]}_{\hbox{\scriptsize N}}$} &
\multicolumn{2}{c|}{$\overline{\delta[\varrho]}_{\hbox{\scriptsize N}}$} &
\multicolumn{2}{c}{$\overline{ s[\varrho]}_{\hbox{\scriptsize N}}$} \\
\hline
1 & 0.666 & 0.149 & 0.554 &0.122 & 0.523 &0.146 & 0.129 & 0.090 
& 0.941 & 0.180 \\  
2 & 0.500 & 0.129 & 0.360 &0.063 & 0.329 &0.077 & 0.194 & 0.077 
& 1.423 & 0.213 \\  
3 & 0.400 &0.107 & 0.264 &0.035 & 0.236 &0.042 & 0.228& 0.066 
& 1.922 & 0.230 \\  
4 & 0.333 &0.089 & 0.208 &0.022 & 0.184 &0.026 & 0.248 & 0.059 
& 2.423 & 0.238 \\  
5 & 0.286 &0.075 & 0.172 &0.015 & 0.151 &0.018 & 0.261 & 0.055 
& 2.928 & 0.245 \\  
6 & 0.250 &0.065 & 0.146 &0.011 & 0.128 &0.013 & 0.269& 0.051 
& 3.431 & 0.251 \\  
7 & 0.222 &0.056 & 0.128 &0.008 & 0.111 &0.010 & 0.275 & 0.049 
& 3.934 & 0.253 \\  
8 & 0.200 &0.049 & 0.113 &0.006 & 0.098 &0.007 & 0.281 & 0.047 
& 4.440 & 0.256 \\  
9 & 0.182 &0.043 & 0.101 &0.005 & 0.088 &0.006 & 0.284 & 0.043 
& 4.944 & 0.260 \\  
10 & 0.166 &0.039 & 0.092 &0.004& 0.080 &0.005 & 0.287 & 0.042 
& 5.447 & 0.261 \\  
11 & 0.154 &0.035 & 0.084 &0.003 & 0.073 &0.004 & 0.290 & 0.041 
& 5.950 & 0.262 \\  
12 & 0.143 &0.032 & 0.078&0.003 & 0.067 &0.004 & 0.292 & 0.039 
& 6.451 & 0.266 \\  
13 & 0.133 &0.029 & 0.072 &0.003 & 0.062 &0.003 & 0.294 & 0.038 
& 6.955 & 0.268 \\  
14 & 0.125 &0.027 & 0.067 &0.002 & 0.058 &0.003 & 0.295 & 0.037 
& 7.456 & 0.269 \\  
15 & 0.118 &0.024 & 0.063 &0.002 & 0.054 &0.002 & 0.297 & 0.037 
& 7.958 & 0.267 \\  
16 & 0.111 &0.023 & 0.059 &0.002 & 0.051 &0.002 & 0.298 & 0.037 
& 8.462 & 0.271 \\  
17 & 0.105 &0.021 & 0.056 &0.002 & 0.048 &0.002 & 0.299 & 0.036 
& 8.964 & 0.271 \\  
18 & 0.100 &0.020 & 0.053 &0.001 & 0.046 &0.002 & 0.300 & 0.035 
& 9.464 & 0.274 \\  
19 & 0.095 &0.018 & 0.050 &0.001 & 0.043 &0.001 & 0.301 & 0.034 
& 9.966 & 0.272 \\  
20 & 0.091 &0.017 & 0.048 &0.001 & 0.041 &0.001 & 0.302 & 0.033 
& 10.467 & 0.273
\end{tabular}}}
\end{table}\\
%%%%%%%%%
As we expected, upon increasing the dimension of the Hilbert
space the average purity decreases and the average of the
symplectic eigenvalue increases. If we rather point our attention on
the average overlap between the random states and their reference
Gaussian states and the average nonG, we observe 
that $\overline{\kappa[\varrho,\tau]}_{\hbox{\scriptsize N}}$ decreases while  
$\overline{\delta[\varrho]}_{\hbox{\scriptsize N}}$ increases.
We also notice 
that, except for the symplectic eigenvalue, the 
variances decrease with the dimension. We conclude that all these 
quantities are concentrating
around typical values.
A more accurate analysis has been made on the distributions of the 
purity $\mu[\varrho]$ and nonG $\delta[\varrho]$. 
%%%%%%%%%%%%%%%%%%%%%%%%%
\begin{figure}[h]
\centerline{
\includegraphics[width=0.24\textwidth]{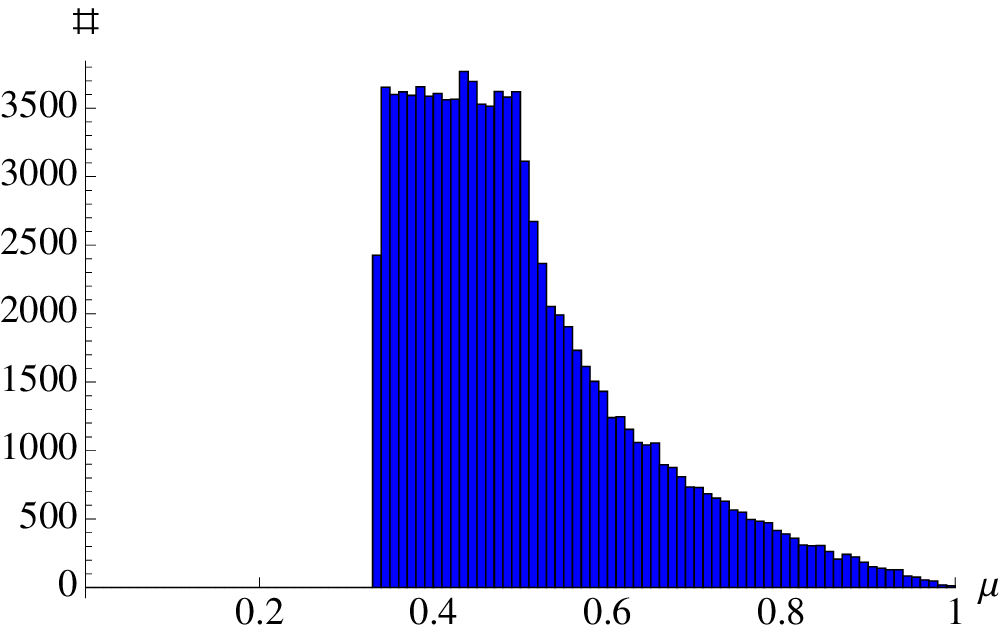}
\includegraphics[width=0.24\textwidth]{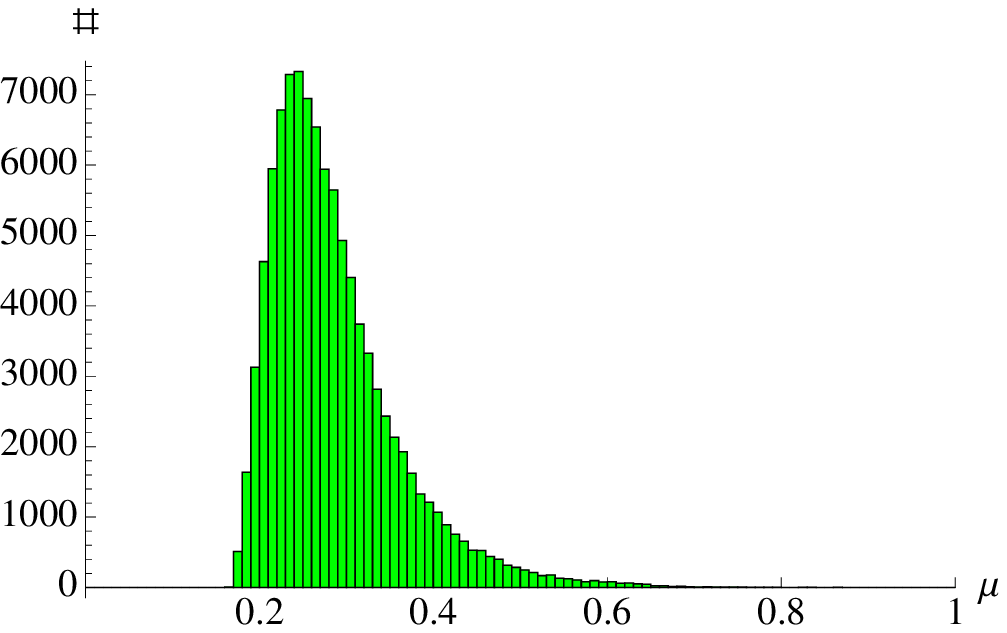}
%}
%\vspace*{4pt}
%\centerline{
\includegraphics[width=0.24\textwidth]{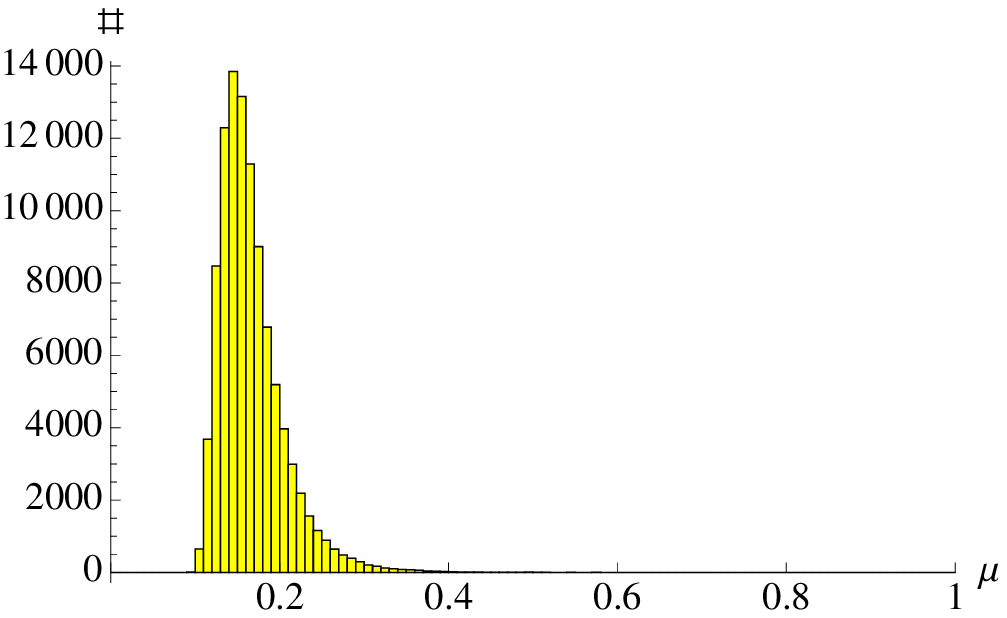}
\includegraphics[width=0.24\textwidth]{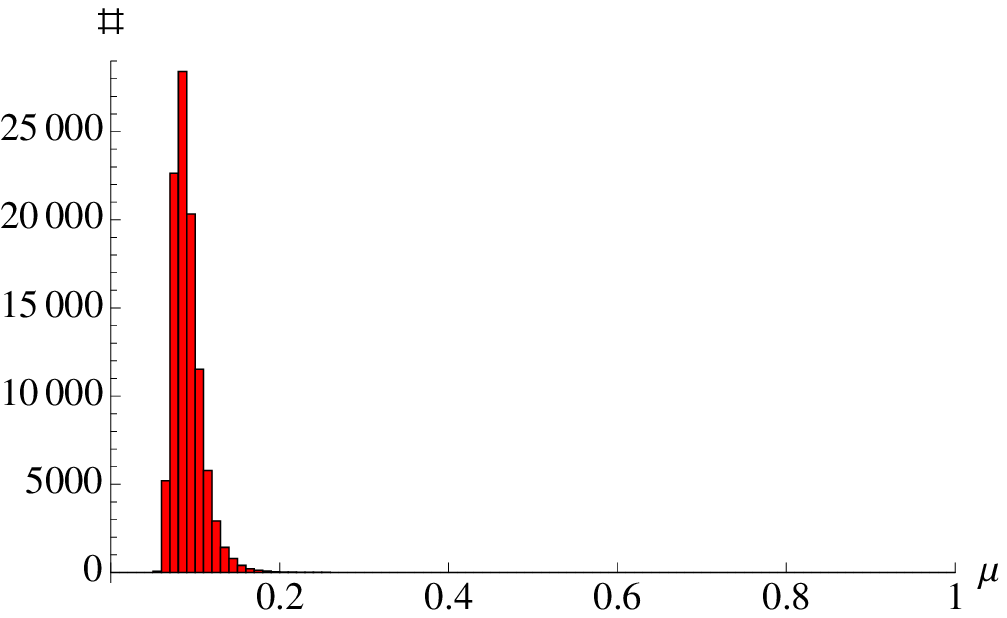}}
%\centerline{
%\includegraphics[width=0.85\textwidth]{HDistMerge.eps}}
%\includegraphics[width=0.45\textwidth]{hist_mu05.eps}}
%\vspace*{4pt}
%\centerline{
%\includegraphics[width=0.45\textwidth]{hist_mu10.eps}
%\includegraphics[width=0.45\textwidth]{hist_mu20.eps}}
%\centerline{\psfig{file=hist_mu02.eps}\psfig{file=hist_mu05.eps} 
%\vspace*{8pt}
\caption{Histograms corresponding to the distributions
of purities of $10^5$ random quantum states for different
dimensions of the Hilbert space. From left to right: 
$N=2$, $N=5$, $N=10$, $N=20$. \label{f:hist_mu}}
\end{figure} \\
%%%%%%%%%%%%%%%%%%%%%%%%%
In Fig. \ref{f:hist_mu} and in Fig. 
\ref{f:hist_d} we show the distributions
of purity and nonG of the $10^5$ random quantum states
for different dimensions of the Hilbert space. We notice that both
these quantities distribute according to Gaussian-like distributions and,
as said before, they concentrate around typical values increasing the 
dimension. 
%%%%%%%%%%%%%%%%%%%%%%%%%
 \\
%%%%%%%%%%%%%%%%%%%%%%%%%%
We also analyzed the behaviour of the average values, in particular
we have looked for fitting functions 
$\overline{\mu}_f=\overline{\mu}_f(N)$
 and
$\overline{\delta}_f=\overline{\delta}_f(N)$
 able to describe the behaviour 
of both $\overline{\mu}_{\hbox{\scriptsize N}}=\overline{\mu[\varrho]}_{\hbox{\scriptsize N}}$ 
and 
$\overline{\delta}_{\hbox{\scriptsize N}}=\overline{\delta[\varrho]}_{\hbox{\scriptsize N}}$
as a function of the maximum number of  photons $N$, 
\emph{i.e.} varying the dimension of the truncated Hilbert space
$d=N+1$. The following fitting functions have been obtained
\begin{align}
\overline{\mu}_f(N) = \frac{2}{N+2} 
\qquad
\overline{\delta}_f(N) = -\frac{1}{(N+2)^{c_1}} + c_2  \label{eq:fit}
\end{align}
wit $c_1=1.560$ and $c_2=0.309$.
These functions
along with the corresponding numerical results are plotted in Fig. \ref{f:fit}.
If we study their behaviour when we consider 
Hilbert spaces with very high dimensions, \emph{i.e.} with maximum number
of photons $N\gg 1$, we observe that the typical purity vanishes as $2/N$ 
while the typical nonG approaches a finite value 
$\overline{\delta}_\infty \approx c_2$. 
%%MG: ho cambiato 
% \overline{\delta}_\infty = c_2 
%in
% \overline{\delta}_\infty \approx c_2
%%%%%%%%%%%%%%%%%%%%%
 \\
%%%%%%%%%%%%%%%%%%%%%
In order to better understand the relationship between the purity and
the nonG of a truncated quantum state, we report the purity and the nonG
of the generated states as points in the plane
$(\mu,\delta)$.  Results are shown in Fig.
\ref{f:pur_nong} where different colors denotes states generated in
Hilbert subspaces with different dimensions.  As it is apparent from the
plot, the points concentrate, at fixed
dimension, in well-defined regions of the plane, whose area decreases
for increasing the dimension. 
Since, as mentioned above, upon increasing the dimension
the typical nonG $\overline{\delta}_{\hbox{\scriptsize N}}$ increases and 
the typical purity $\overline{\mu}_{\hbox{\scriptsize N}}$ decreases, we have 
that higher typical nonG corresponds to lower typical purities.
On the other hand, if we rather focus to points 
fixed dimension, we have that large values of nonG correspond 
to large values of purity, this effect being more pronounced for higher 
values of $N$.
%%%%%%%%%%%%%%%%%%%%%%
 \\
%%%%%%%%%%%%%%%%%%%%%%
Using Eqs. (\ref{eq:fit}) we may write the relation between the 
typical nonG and the typical purity as  
$$ 
\overline{\delta}_f(\overline{\mu}) 
= c_2 - \left( \frac{\overline{\mu}}{2}\right)^{c_1}\:.$$
Comparison with numerical findings is reported in the right panel
of Fig. \ref{f:pur_nong}.
\section{Conclusion}\label{s:out}
In conclusion, we have analyzed the properties of random quantum states
generated in finite dimensional Hilbert spaces in terms of their nonG
character and purity. We have found that both quantities distribute 
according to a Gaussian-like distribution with variance that decreases
by increasing the dimension, \emph{i.e.} they concentrate around typical values.
We also found that the typical nonG and the typical purity are monotone
functions of the dimension $d$. In particular, the average purity decreases 
to zero whereas the average nonG increases to an asymptotic value.
Besides, we have found that, at fixed dimension, the points
corresponding to the random states in the plane $(\mu,\delta)$ 
are confined in well-defined regions whose area decreases with the
dimension. For increasing dimension higher nonG correspond to higher 
purities.
%%%%%%%
\section*{Acknowledgments}\label{s:ackn}
We thank Konrad Banaszek for several discussions about non-Gaussianity.
This work has been partially supported by the CNR-CNISM convention.
%%%%%%%

%%%% 

\begin{thebibliography}{99}
\bibitem{r1} P. Walther et al., Phys. Rev. A {\bf 75}, 012313 (2007). 
\bibitem{r2} B.M. Escher et al., Phys. Rev. {\bf 72}, 045803  (2005)
\bibitem{q1} L. A. de Souza et al., Phys. Lett. A {\bf 309}, 5 (2003);
\bibitem{t1} A.T. Avelar et al., Phys. Lett. A  {\bf 318}, 161 (2003).  
\bibitem{zyc01} K. Zyczkowski et al., Phys. Rev. A {\bf 58}, 883 (1998).
\bibitem{zyc02} K. Zyczkowski, Phys. Rev. A {\bf 60}, 3496 (1999).
\bibitem{cap01} V. Cappellini et al., Phys. Rev. A {\bf 74}, 062322 (2006).
\bibitem{GRG}
J. Eisert et al., Int. J. Quant. Inf. {\bf 1}, 479 (2003);
A. Ferraro et al., {\em Gaussian States in Quantum Information},
(Bibliopolis, Napoli, 2005); F. Dell'Anno et al., Phys. Rep. {\bf
428}, 53 (2006).
\bibitem{Tom} T. Opatrny et al., Phys.Rev. A {\bf
61}, 032302 (2000).
\bibitem{IPS2a} P. T. Cochrane et al., Phys Rev. A {\bf 65}, 062306 (2002).
\bibitem{IPS2b} S. Olivares et al., Phys. Rev. A {\bf 67}, 032314
(2003).
\bibitem{nonGclon} N. J. Cerf et al., Phys. Rev. Lett. {\bf 95}, 070501 (2005).
\bibitem{IPS1} S. Olivares and M. G. A. Paris, J. Opt. B {\bf 7}, S392 (2005).
\bibitem{KorolkovaKerr} T. Tyc at al., New. J. Phys. {\bf 10},  (2008).
\bibitem{IPS_Wenger} J. Wenger et al,
Phys. Rev. Lett. {\bf 92}, 153601 (2004).
\bibitem{nonG_HS} M. G. Genoni et al., Phys. Rev. A {\bf 76}, 042327 (2007).
\bibitem{zyc03} K. Zyczkowski and M. Kus, J. Phys. A: Math. Gen. {\bf 27}, 4235-4245 (1994).
\bibitem{Glauber2} K. E. Cahill and R. J. Glauber, Phys. Rev. {\bf 177}, 1882--1902 (1969).
\bibitem{Marian} P. Marian et al., Phys. Rev. A {\bf 47}, 4474 (1993); 4487 (1993).
%%% 
\end{thebibliography}
\end{document}